\documentclass{article}

\usepackage[utf8]{inputenc} 
\usepackage[T1]{fontenc}    
\usepackage{hyperref}       
\usepackage{url}            
\usepackage{booktabs}       
\usepackage{amsfonts}       
\usepackage{amssymb}
\usepackage{amsmath}
\usepackage{nicefrac}       
\usepackage{microtype}      
\usepackage{xcolor}         
\usepackage{cleveref}       
\usepackage[a4paper, total={6in, 8in}]{geometry}
\usepackage{graphicx}
\usepackage[round]{natbib}

\newcommand{\G}{\mathcal{G}}
\newcommand{\setX}{\mathcal{X}}

\title{Equivariant Bootstrapping for Uncertainty Quantification in Imaging Inverse Problems}
\author{Julián Tachella$^{1}$ and Marcelo Pereyra$^{2}$}

\date{%
    $^1$CNRS \& ENS Lyon\\%
    $^2$Heriot-Watt University\\[2ex]%
    \today
}
\begin{document}

%

%

\maketitle

\begin{abstract}
Scientific imaging problems are often severely ill-posed, and hence have significant intrinsic uncertainty. Accurately quantifying the uncertainty in the solutions to such problems is therefore critical for the rigorous interpretation of experimental results as well as for reliably using the reconstructed images as scientific evidence. Unfortunately, existing imaging methods are unable to quantify the uncertainty in the reconstructed images in a manner that is robust to experiment replications. This paper presents a new uncertainty quantification methodology based on an equivariant formulation of the parametric bootstrap algorithm that leverages symmetries and invariance properties commonly encountered in imaging problems. Additionally, the proposed methodology is general and can be easily applied with any image reconstruction technique, including unsupervised training strategies that can be trained from observed data alone, thus enabling uncertainty quantification in situations where there is no ground truth data available. We demonstrate the proposed approach with a series of numerical experiments and through comparisons with alternative uncertainty quantification strategies from the state-of-the-art, such as Bayesian strategies involving score-based diffusion models and Langevin samplers. In all our experiments, the proposed method delivers remarkably accurate high-dimensional confidence regions and outperforms the competing approaches in terms of estimation accuracy, uncertainty quantification accuracy, and computing time\footnote{The code associated with this paper can be found at~\url{https://github.com/tachella/equivariant_bootstrap}.}.
\end{abstract}

\section{Introduction}
Digital images inform decisions that have a major impact on the economy, society, and the environment (e.g., in medicine, agriculture, forestry, astronomy, and defense). Early uses of images as evidence in decision-marking or science were predominantly qualitative. However, modern strategies increasingly adopt a so-called quantitative imaging approach which recognises images as high-dimensional physical measurements and seeks to use the images as quantitative evidence. The approach is particularly prominent in scientific applications. 

Performing inference in imaging problems usually requires solving a high-dimensional inverse problem that is severely ill-posed. In particular, many commonly encountered tasks related to image restoration and reconstruction involve performing inference on an unknown image $x_\star$ taking values in a signal set $\setX \subset \mathbb{R}^n$, from a single measurement $y \in \mathbb{R}^m$ modeled as a realization of
\begin{equation}
\label{eq: sampling_distribution}
    Y \sim P(A x_\star)
\end{equation}
where $P$ is a statistical model describing stochastic aspects of the data acquisition process (e.g., measurement noise) and $A \in \mathbb{R}^{m \times n}$ describes deterministic instrumental aspects of the observation process. For example, such models are frequently encountered in problems related to image deblurring, inpainting, super-resolution, compressive sensing reconstruction, and tomographic reconstruction (see, e.g., \citep{Kaipio2005,Ongie2020}). Unfortunately, the models that accurately describe the data acquisition process are usually either weakly or not identifiable, and hence there is significantly uncertainty about $x_\star$ after observing $y$.

State-of-the-art imaging methods address this difficulty by intimately combining computational imaging and machine learning techniques, which learn to recover $x_\star$ from $y$ by leveraging information from large training datasets \citep{Ongie2020,Mukherjee2023}. Early learning-based imaging methods required abundant ground truth data for training. However, new unsupervised strategies exploit symmetries or equivariance properties in the problem to learn estimators of $x_\star$ from the observed data alone \citep{chen2023imaging}.

Despite sustained progress in image estimation accuracy, important aspects of quantitative imaging sciences are still in their infancy. In particular, most imaging methods cannot reliably quantify the uncertainty in the delivered solutions, which is essential for the rigorous interpretation of experiments and robust interfacing of imaging pipelines with decision-making processes. Furthermore, as we illustrate through numerical experiments in Section 5, even the most powerful uncertainty quantification (UQ) methods currently available cannot meaningfully quantify the uncertainty in the reconstructed images in a manner that is robust to frequentist validation and experiment replications. This critical limitation hinders the value of images as quantitative evidence for decision-making and science.

This paper presents the following main contributions: 
\begin{enumerate}
    \item We propose the equivariant bootstrap algorithm, a new parametric bootstrapping technique that exploits symmetries in the problem in order to construct accurate confidence regions for $x_\star$, even in situations where the model is not identifiable.
    
    \item We present a theoretical analysis of the equivariant bootstrap for the case of a linear estimator, which provides clear insights into the key factors driving the accuracy of the method.
    
    \item We demonstrate the effectiveness of the proposed equivariant bootstrap through a series of experiments involving three different imaging inverse problems, where the method consistently outperforms alternative strategies from the state-of-the-art in terms of estimation accuracy, uncertainty quantification accuracy, and computing time.
\end{enumerate}

\begin{figure*}[t]
    \centering
    \includegraphics[width=1\textwidth]{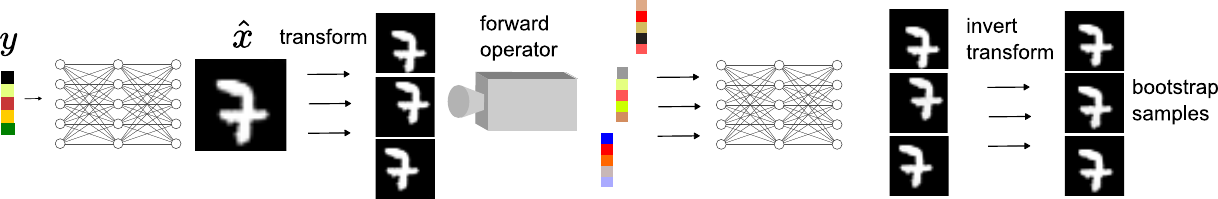} 
    \caption{\textbf{Illustration of the equivariant bootstrap method.} Additional measurement vectors are bootstrapped by applying random transformations such as translations and/or rotations, which are then used to quantify the error associated with the reconstruction network.} \label{fig: schematic}
\end{figure*}


\section{Related Work}
\paragraph{Bayesian Methods}
Modern imaging methods rely strongly on the Bayesian statistical framework to perform UQ tasks. In particular, Bayesian imaging strategies with data-driven priors encoded by neural networks have received a lot of attention lately. Within this context, state-of-the-art methods involve the prior through its score function, which can be related to an image denoising network via Tweedie's formula (see, e.g., plug-and-play ULA ~\citep{laumont2022bayesian}, and the denoising diffusion methods DPS~\citep{chung2022diffusion}, DDRM~\citep{kawar2022denoising}, and diffPIR~\citep{zhu2023denoising}). Alternatively, many modern methods rely on deep generative architectures; e.g., variational autoencoders \citep{Holden2022,Zhang2021}, generative adversarial networks \citep{arridge_maass_öktem_schönlieb_2019}, and normalizing flows \citep{Hertrich2023}. These can be used to encode the prior within a Bayesian model or to approximate the posterior distribution directly. Otherwise, some methods perform UQ via Laplace approximations (e.g., \citep{antorán2022uncertainty}).

\paragraph{Stein's Unbiased Risk Estimator (SURE)}
SURE is a highly effective approach for estimating the mean squared error (MSE) in denoising problems~\citep{stein1981estimation}. The method can be extended to other inverse problems~\citep{eldar2008generalized}, but it can only reliably quantify the error in the range space of the forward operator. As a result, it is not directly useful for UQ in inverse problems with a non-trivial nullspace.

\paragraph{Conformal Prediction}
Conformal prediction is a model-free UQ approach that leverages a calibration dataset and exchangeability in order to build confidence intervals that are marginally exact (i.e., probabilities are computed w.r.t. the joint distribution of the unknown image and the measurement data, without conditioning of the observed measurement data). Several recent papers propose conformal methods for imaging problems (see, e.g., ~\cite{angelopoulos2022conformal}), focusing predominantly on UQ tasks at the scale of a single pixel. 
Scaling conformal prediction to large image structures requires careful construction of the conformity function that determines the shape of the confidence region, a difficulty that is usually addressed by using the regions provided by a different UQ technique and applying a conformal calibration step (see, e.g., \citep{narnhofer2022posteriorvariancebased}). The method proposed in this paper could be easily combined with a conformal correction step if a calibration set is available.

\paragraph{Bootstrap Methods}
Bootstrap resampling is a powerful general strategy for assessing the variability of estimators and other statistics w.r.t. the data sampling distribution ~\citep{efron1994introduction}. Although widely used for UQ in other disciplines, the bootstrap is difficult to apply directly to imaging problems, as these are often severely ill-posed because of identifiability issues in the likelihood function. The method proposed in this paper builds on the parametric formulation of the bootstrap, which we adapt in order to tackle models that are weakly or not identifiable (i.e., with a forward operator $A$ that is not invertible).

\paragraph{Equivariant Imaging}
Equivariance plays an important role in imaging inverse problems, enabling unsupervised learning~\citep{chen2021equivariant, chen2021robust} and improving the generalization of the estimators~\citep{mohan2019robust, celledoni2021equivariant} (e.g., see the recent tutorial~\cite{chen2023imaging}). The proposed bootstrap endows the equivariant imaging approach~\citep{chen2021equivariant, chen2021robust} with powerful UQ capabilities.

\section{Parametric Bootstrap}\label{sec: parametric bootstrap}
The standard parametric bootstrap method probes the accuracy of $\hat{x}(y)$ by constructing an i.i.d sample $\{\tilde{x}^{(1)},\ldots,\tilde{x}^{(n)}\}$ which approximates the sampling distribution of $\hat{x}(Y)$ when $Y \sim P(A x_\star)$. The sample is constructed as follows: for any $i = \{1,\ldots,N\}$
\begin{enumerate}
\item Given $\hat{x}(y)$, draw a bootstrap measurement $\tilde{y}^{(i)}$ from the sampling distribution
    $$
    \tilde{Y}^{(i)} \sim P(A \hat{x}(y))\,.
    $$
\item Given the bootstrap measurement $\tilde{y}^{(i)}$, compute
$$
\tilde{x}^{(i)}=\hat{x}(\tilde{y}^{(i)})\,.
$$
\end{enumerate}

The realized sample $\{\tilde{x}^{(1)},\ldots,\tilde{x}^{(n)}\}$ can then be used, for example, to construct confidence regions for $x_\star$. In our experiments, we consider the confidence region
\begin{equation}
\mathcal{C}_\alpha = \{x \in \setX : \|x-\hat{x}(y)\|_2^2 < q_{\alpha}\}\,,
\end{equation}
with $q_{\alpha}$ set to the top $\alpha$-quantile of the sample $\{\|\tilde{x}^{(i)}-\hat{x}(y)\|^2\}_{i=1}^{n}$. One can also design more compact and geometrically interesting regions by using the sample covariance of $\{\tilde{x}^{(1)},\ldots,\tilde{x}^{(n)}\}$. The sample is also useful for exploring the accuracy of $\hat{x}(Y)$ (e.g., bias, variance). 

Although widely used in other settings, the parametric bootstrap often performs poorly in imaging problems. In particular, it severely underestimates the uncertainty in the solution because of the bias arising from using an estimate $\hat{x}(y)$ of $x_\star$ in the sampling distribution of $Y$, as opposed to the true value $x_\star$ which generated $y$. 

For illustration, consider a severely under-determined problem of the form $y = A x_\star + e$ where $m \ll n$ and $e$ is a realization of Gaussian noise. In such problems, there is significant uncertainty about the solution on the nullspace of $A$. For simplicity, consider the linear estimator $\hat{x}(y) = My$ for all $y \in \mathbb{R}^m$, that seeks to invert $A$ on $\setX$. Suppose that we use the conventional parametric bootstrap to estimate the MSE $\|\hat{x}(y)-x_\star\|_2^2$. For the considered setup, we have that the true MSE is given by
\begin{equation}
    \|\hat{x}(y)-x_\star\|_2^2 = \|MAx_\star + Me - x_\star\|_2^2\,,
    \label{trueMSE_noise}
\end{equation}
which is approximately 
\begin{equation}
    \|\hat{x}(y)-x_\star\|_2^2 \approx \|(MA - I)x_\star\|_2^2\,,
\end{equation}
when the error stemming from the noise $e$ is negligible relative to the error from incorrectly inverting $A$ on $\setX$. Conversely, in that same scenario, the parametric bootstrap produces the estimate 
\begin{equation}
    \textrm{E}_{\tilde{Y}}\|\hat{x}(\tilde{Y})-\hat{x}(y)\|_2^2 \approx \|(MA-I_n)MAx_\star\|_2^2\,,
    \label{parametric_bootstrap_MSE}
\end{equation}
which can be potentially much smaller than the true error $\|(MA - I)x_\star\|_2^2$. For example, if $MA$ is a projection matrix, the bootstrap error $\|(MA-I_n)MAx_\star\|_2^2 = 0$ regardless of the true error. This drawback of the parametric bootstrap is illustrated through a series of experiments in Section 4, where we consistently observe an underestimation of the uncertainty in the solution. 


\section{Equivariant Bootstrap}
\subsection{Proposed method} \label{subsec: proposed method}
The proposed equivariant bootstrap method is a parametric bootstrapping technique designed for situations in which the set of signals $\mathcal{X}$ is known to be invariant to a certain group of transformations. The aim is to leverage these symmetries in order to mitigate the bias arising from using an estimate $\hat{x}(y)$ of $x_\star$ in the approximation of the sampling distribution of $Y$. The method is particularly useful in situations where access to ground truth data is difficult or not possible, and it is necessary to quantify the uncertainty in the delivered solutions from the observed measurement data alone.

Let $\G = \{g_1,\ldots,g_{|\G|}\}$ be a finite group acting on $\setX$, whose action is represented by the invertible linear mappings $T_g$. We assume that $\setX$ is $\G$-invariant; i.e., for all $x \in \setX$ and all $g \in \G$, we have that 
$$
T_g x \in \setX\,.
$$
In a manner akin to the conventional parametric bootstrap, the proposed equivariant bootstrap method probes the accuracy of $\hat{x}(y)$ by constructing an i.i.d sample $\{\tilde{x}^{(1)},\ldots,\tilde{x}^{(n)}\}$ which approximates the sampling distribution of $\hat{x}(Y)$ when $Y \sim P(A x_\star)$. The sample is constructed as follows: for any $i = \{1,\ldots,N\}$
\begin{enumerate}
    \item Draw a random transform $g_i$ uniformly from $\G$.
    \item Given $g_i$ and $\hat{x}(y)$, draw a bootstrap measurement $\tilde{y}^{(i)}$ from the sampling distribution
    $$
    \tilde{Y}^{(i)} \sim P(AT_{g_i} \hat{x}(y))\,.
    $$
\item Given the generated bootstrap measurement $\tilde{y}^{(i)}$ and $g_i$, compute
$$
\tilde{x}^{(i)}=T_{g_i}^{-1}\hat{x}(\tilde{y}^{(i)})\,.
$$
where $T_{g_i}^{-1}$ inverts the action $g_i$. 
\end{enumerate}

It is useful to view the equivariant bootstrap procedure as a data augmentation strategy that constructs an augmented sampling distribution 
$$
G \sim \mathcal{U}(\G),\quad (Y|G=g,x) \sim P(AT_gx)\,,
$$
where $G$ is an auxiliary variable that takes values uniformly in $\G$. Without loss of generality, we view the observed data $y$ is a realization of this augmented model with $g = 1_{\G}$, the identity element of $G$.

As mentioned previously, in many imaging inverse problems the uncertainty about the solution stems predominantly from the fact that $A$ has a large nullspace, which leads to significant non-identifiability issues in the likelihood function. 
The equivariant bootstrap exploits the fact that when $A$ is not $\G$-equivariant, the composition $AT_g$ can span a different subspace than $A$, such that the operator resulting from averaging $AT_g$ over $\G$ can be full rank. The equivariant bootstrap leverages this property to probe the variability of the estimator $\hat{x}(Y)$ and characterize the uncertainty in $x_\star$. 

\subsection{Analysis With a Linear Estimator} \label{subsec: analysis}
In order to develop an intuition for the equivariant bootstrap and the effect of introducing the actions $\G$, we consider a noise-free problem $y = Ax_\star$ with $m \ll n$, where there is significant uncertainty about the solution because $A$ has non-trivial nullspace. For simplicity, we consider again a linear estimator $\hat{x}(y) = My$ for all $y \in \mathbb{R}^m$ that seeks to invert $A$ on $\setX$, and use the proposed equivariant bootstrap to estimate the MSE $\|\hat{x}(y)-x_\star\|_2^2$. We assume that $\setX$ is a low-dimensional subspace that is $\G$-invariant for some compact group $\G$. For example, $\G$ could be the group of cyclic shifts associated with circulant matrices $T_{g}$ which shift the image by $g$ pixels, with $g \in \{0, \ldots, n-1\}$ (see \cite[Section 4.1]{tachella2022sensing} for other commonly encountered examples). In practice, $\setX$ is unknown so the estimator $M$ will fail to perfectly invert $A$. We let $B = MA$, and use the decomposition $B = B_\star + R$, where $B_\star$ is a $\G$-equivariant orthogonal projection onto $\setX$, and $R = B - B_\star$ is a residual term representing the estimator error. We will see that the equivariant bootstrap is effective when $R$ is small but also far from being $\G$-equivariant. In that case, the actions of $\G$ will ``average out'' the bias introduced by $R$ and improve the quality of the bootstrap sample as a result. 

We now introduce some elements of linear representation theory that are essential to our analysis \citep{serre1977linear}. For any compact group $\G$, $T_g$ with $g \in \G$ admits a linear representation $T_g = F^{-1} \Lambda_{g} F$ where $\Lambda_{g}$ is a block-diagonal matrix and $F$ is an orthonormal basis on $\mathbb{C}^{n}$ associated with $\G$, but independent of $g$. For example, when $\G$ represents the group of cyclic shifts, we have that $F$ is the discrete Fourier transform and $\Lambda_{g}$ is a diagonal matrix containing the Fourier transform of the discrete shift operation (see \cite[Section 4.1]{tachella2022sensing} for more details and other examples). More generally, any $\G$-equivariant matrix $C \in \mathbb{R}^{n\times n}$ also admits a linear representation $C = F^{-1} \Lambda_{C} F$, where $\Lambda_{C}$ is a block-diagonal with the same structure or support as $\Lambda_g$. It follows that any $\G$-equivariant matrix $C$ commutes with $T_g$ for any $g \in \G$. Conversely, $R$ is not $\G$-equivariant, and therefore $F R F^{-1}$ is usually a dense matrix that does not commute with $T_g$ for any $g \in \G$.

We analyze the estimation by equivariant bootstrapping of the MSE between $\hat{x}(y)-x_\star$, which also underpins $\mathcal{C}_\alpha$. For the considered setup, 
\begin{equation}
    \label{eq:trueMSE}
    \|\hat{x}(y)-x_\star\|^2 = \|B x_\star - x_\star\|^2 = \|Rx_\star\|^2\,.
\end{equation}

The equivariant bootstrap estimate of \eqref{eq:trueMSE} is given by

\begin{align*}
\mathbb{E}_{\hat{Y}} \|\tilde{x}(\tilde{Y})-Bx_\star\|_2^2 &= \frac{1}{{|\G|}}\sum_{g\in\G} \|T^{-1}_g B T_g B x_\star - Bx_\star\|_2^2  \\
&= \frac{1}{{|\G|}}\sum_{g\in\G} x_\star^\top B^\top (I_n + T^{-1}_g B^\top B T_g - 2 T^{-1}_g B T_g)Bx_\star \\
&= x_\star^\top B^\top \left(I_n + \Pi_{\G}(B^\top B) - 2 \Pi_{\G}(B)\right)Bx_\star ,
\end{align*}
where for any matrix $C \in \mathbb{R}^{n \times n}$, $\Pi_{\G}(C) \triangleq \sum_g T^{-1}_g C T_g / |\G|$ denotes the so-called Reynolds averaging operator associated with $\G$ \citep{serre1977linear}. $\Pi_{\G}$ is a projection operator onto the linear subspace of matrices that are diagonal on the basis $F$ associated with $\G$. Consequently, $\Pi_{\G}(C) = C$ for any matrix $C \in \mathbb{R}^{n \times n}$ that is $\G$-equivariant. Note that if $C$ is not $\G$-equivariant, $\Pi_{\G}$ will potentially significantly shrink the Frobenius norm of $C$ by setting all its off-block-diagonal elements (in the basis $F$) to zero.

By using the decomposition $B = B_\star + R$ introduced previously and that $\Pi_{\G}$ is a linear operator, we obtain 

\begin{equation*}
\begin{split}
\mathbb{E}_{\hat{Y}}&\|\tilde{x}(\tilde{Y})-Bx_\star\|_2^2  = \underbrace{\| Rx_\star \|^2}_{\text{true error}} - \underbrace{\| Rx_\star \|_{B_{\star}}^2}_{\text{bias term 1}}  + \underbrace{x_\star^\top B^\top  \Pi_{\G}\left( R^\top R + R^{\top}B_\star +B_\star R -2 R \right)Bx_\star}_{\text{bias term 2}} \\ 
\end{split}   
\end{equation*}
where we have also used the that $B_\star$ is the orthogonal projection onto $\setX$ and therefore $B_\star x_\star = x_\star$, $B_\star^\top = B_\star$ and $\Pi_{\G}(B_\star) = B_\star$. 

We now analyze the two bias terms which can lead to a biased estimation of the true error. If the estimator is measurement consistent, i.e., if it verifies $A\hat{x}(y)=y$, then the first bias term is zero, see \Cref{app: proofs} for a detailed derivation. This is a simple property that most estimators verify in practice.

The second bias term depends on whether the matrix $R$ is $\G$-equivariant or not. Due to the averaging operator, we can rewrite this second term as 
$$
x_\star^\top B^\top F^{-1} \Lambda F Bx_\star, 
$$
where $F$ is an orthonormal basis and $\Lambda$ is a block-diagonal matrix with 
$$
0 \leq \| \Lambda \|_{\textrm{Frob}} \leq \| R^\top R + R^{\top}B_\star +B_\star R -2 R\|_{\textrm{Frob}}
.$$
If the error term $R$ is not $\G$-equivariant (and thus dense on the basis $F$), then the norm of $\Lambda$ will be much smaller than the norm of $R$, and consequently, the second bias term will be small. Conversely, if $R$ is $\G$-equivariant, $\Lambda$ will have a  norm similar to $R$ and the averaging will not mitigate the second bias term.

We conclude that the equivariant bootstrap can potentially significantly reduce the bias inherent in parametric bootstrapping, especially when the estimator is measurement consistent, the signal $\setX$ is $\G$-invariant and low-dimensional and either the forward operator and/or the estimator are not $\G$ equivariant. 
Natural signal sets often exhibit a range of symmetries (e.g., invariance to rotation, translations, permutations, etc. \citep{tachella2022sensing}). The key to leveraging these symmetries in order to reduce the bootstrap bias is to identify a symmetry group $\G$ that ``averages out'' the estimation error while leaving the signal set $\setX$ unchanged. This arises when $\setX$ is $\G$-invariant but the estimation error is far from $\G$-equivariance.

\section{Experimental Results}\label{sec: experiments}
We now present a series of numerical experiments and comparisons with alternative statistical UQ imaging approaches from the state of the art. We use each method to compute a confidence region for $x_{\star}$, derived from the pivotal statistic $\|x_{\star}-\hat{x}(Y)\|_2^2$ related to the estimation MSE. We evaluate the accuracy of these confidence regions by calculating the empirical coverage probabilities on a test set, as measured by the proportion of test images that lie within the confidence regions for a range of specified confidence levels between 0\% and 100\%. We perform three types of experiments:  compressed sensing image reconstruction, sparse-angle tomography, and image inpainting. All our experiments were performed using the \texttt{deepinv} open-source library~\citep{deepinverse} on a local cluster with 4 NVIDIA RTX 3090 GPUs.
 
\paragraph{Compressed Sensing}
We use the MNIST dataset, which consists of images of $28\times 28$ pixels. We use $6\times10^{4}$ images for training, and $384$ images for testing. The forward operator is defined using $m=256$ measurements with entries sampled from a Gaussian distribution with zero mean and variance $1/m$. Measurements are corrupted with Gaussian noise of standard deviation $0.05$.

\paragraph{Inpainting}
The DIV2K dataset~\citep{Agustsson_2017_CVPR_Workshops} contains 2000 high-resolution RGB images. We generate a dataset of $2\times 10^{4}$ crops of $256\times 256$ pixels for training and $200$ for testing. The measurement data is obtained by applying an inpainting mask with binary entries sampled from a Bernoulli distribution with a probability of 0.5 and adding white Gaussian noise with a standard deviation of $0.05$.

\paragraph{Sparse-Angle Tomography}
The LIDC-IDRI dataset~\citep{armato2011lung} consists of computed tomography scans from 1010 patients. We re-scale the 2D slices to a size of $256\times 256$ pixels and use the central slices of $100$ patients for testing and $5000$ slices from the remaining patients for training. The measurement data are obtained by using 40 projections, taken at equally distanced angles, corrupted with Gaussian noise of standard deviation $0.1$.

\begin{table*}[t]
\caption{Average test PSNR in dB for the evaluated methods. } \label{tab: psnr}
\begin{tabular}{l|ccccc}
 & \begin{tabular}[c]{@{}c@{}}Diffusion\\ (DDRM)\end{tabular} & \begin{tabular}[c]{@{}c@{}}Diffusion\\ (diffPIR)\end{tabular} & ULA & \begin{tabular}[c]{@{}c@{}}Proposed bstrap \\ (unsup. model)\end{tabular} & \begin{tabular}[c]{@{}c@{}}Proposed bstrap\\ (sup. model)\end{tabular} \\ \hline
\begin{tabular}[c]{@{}l@{}}C. Sensing\\ (MNIST)\end{tabular} & - & -
& $28.54\pm2.25$ & $\boldsymbol{34.11\pm2.09}$ & $33.9\pm2.32$ \\
\begin{tabular}[c]{@{}l@{}}Inpainting\\ (DIV2K)\end{tabular} & $32.27\pm3.95$ & $30.51\pm3.74$ & $30.52\pm3.35$ & $31.56\pm4.12$ & $\boldsymbol{32.47\pm3.87}$ \\
\begin{tabular}[c]{@{}l@{}}Tomography\\ (LIDC)\end{tabular} & - & 
$37.02\pm 0.79$ & $35.85\pm0.54$ & $37.38\pm 0.65$ & $\boldsymbol{41.03\pm0.91}$
\end{tabular}
\end{table*}
Moreover, we report comparisons between our method and the following techniques from the state of the art:
\begin{enumerate}
    \item \textbf{Diffusion methods:} We evaluate two recently proposed score-based diffusion models, namely DDRM~\citep{kawar2022denoising} and diffPIR~\cite{zhu2023denoising}. These are Bayesian posterior sampling strategies that rely on pre-trained denoising score-matching networks that encode the prior information available. We use the denoiser network in~\citep{choi2021ilvr} for RGB images and the one in~\citep{zhang2017learning} for single-channel images. By running the diffusion model multiple times we obtain a set of Monte Carlo samples from the posterior distribution associated with each method, which we then use to approximate the posterior mean $\hat{x}(y)=\mathbb{E}\{x|y\}$ as well as a posterior credible region based on an $\ell_2$-ball around $\hat{x}(y)$, derived from the posterior distribution of the error $\|x-\hat{x}(y)\|_2^2$. 
  
    \item \textbf{Unadjusted Langavin Algorithm:} We consider the plug-and-play Unadjusted Langevin Algorithm (ULA) introduced by~\cite{laumont2022bayesian}. This Bayesian posterior sampling strategy implements a Langevin MCMC scheme with a pre-trained MMSE denoiser in lieu of the gradient of the log-prior density. For stability, the denoiser is trained with a controlled Lipschitz constant. We use the stable DnCNN denoiser introduced in~\citep{terris2020building}. As with the diffusion methods, we use the Monte Carlo samples generated by ULA to approximate $\hat{x}(y)=\mathbb{E}\{x|y\}$ and to compute a posterior credible region based on an $\ell_2$-ball around $\hat{x}(y)$.
    
    \item \textbf{Naive bootstrap:} For completeness, we also evaluate the conventional parametric bootstrap introduced in Section 3, based on the sampling distribution $\tilde{Y} \sim P(A\hat{x}(y))$, without using a group action. We use the bootstrap sample to determine the sampling distribution of the error $\|\hat{x}(\tilde{Y})-\hat{x}(y)\|_2^2$. The only source of randomness in this case is the stochasticity associated with $P$ (e.g., measurement noise).
    
    \item \textbf{Equivariant bootstrap:} We evaluate the proposed method on estimator networks $\hat{x}(y)$ trained to minimize the reconstruction MSE. We consider both standard supervised training using a dataset pairs $\{(x_i,y_i)\}_i$, as well as a unsupervised training using the method in~\cite{chen2021robust} which only requires a dataset of measurement vectors $\{y_i\}_i$. We consider two forms of group actions: rotations and two-dimensional shifts. Rotations are sampled from a Gaussian distribution with zero mean and standard deviation of $\sigma_{\theta}$, while horizontal and vertical shifts are sampled from a uniform distribution on $[-\Delta t, \Delta t]$ pixels. In practice, we compute the MSE samples as $\| T_{g_i}\hat{x}(y) - \hat{x}(\tilde{y}^{(i)})\|^2$, avoiding\footnote{This bypasses interpolation or border artifacts associated with rotations that are not multiple of 90 degrees.} the use of the inverse transform $T_{g_i}^{-1}$.
\end{enumerate}

\begin{figure*}[t]
    \centering
    \includegraphics[width=.95\textwidth]{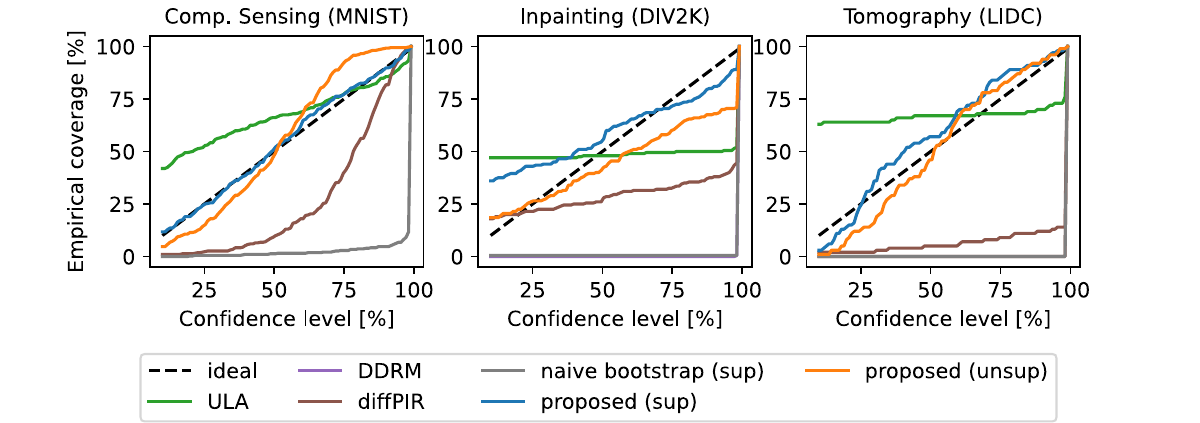} 
    \caption{Coverage plots for all the evaluated methods on 3 inverse problems. Methods with a coverage closer to the dotted line provide a more precise quantification of the uncertainty of the estimates.} \label{fig: all coverages}
\end{figure*}
All methods are used to generate 100 independent Monte Carlo samples, from which we compute the confidence regions for various confidence levels (in the case of PnP-ULA we achieve quasi-independent samples by applying thinning). Using a larger number of Monte Calo samples was not possible because of the large number of repetitions and deep neural function evaluations (NFEs) required by the diffusion models. \Cref{table: nfes} compares the computational load of the evaluated methods in terms of the number of NFEs. Diffusion-based methods require 100 NFEs per sample, and they are the most computationally demanding method across the evaluated algorithms. ULA requires a single NFE per MCMC iteration, however, we use a 1-in-30 thinning factor to reduce the correlation in the chain, so the cost is approximately 30 NFEs per sample. Remarkably, the bootstrap method requires only a single NFE per sample and thus can obtain uncertainty estimates in the order of seconds even for large images. 

\begin{table}[h]
\centering
\caption{Neural function evaluations (NFEs) per Monte Carlo (MC) sample. }
\label{table: nfes}
\begin{tabular}{l|ccc}
\textbf{Method}             & \textbf{Diffusion} & \textbf{ULA} & \textbf{Bootstrap} \\ \hline
NFEs/MC sample & 100       & 30  & 1       
\end{tabular}
\end{table}

\Cref{tab: psnr} shows the peak signal-to-noise ratio (PSNR) for each evaluated method and each imaging problem considered, while \Cref{fig: all coverages} shows the coverage probabilities of their delivered confidence regions. Examples of measurements and reconstructed images are included in \Cref{app: additional results}. For these experiments, the hyper-parameters of all the methods (bootstrap parameters in~\Cref{tab: bootstrap params}, regularisation parameter in ULA, etc.), were calibrated on a small evaluation set of 16 images per problem. 

Notice that while most competing methods produce relatively accurate UQ results for the moderately low-dimensional images of the MNIST dataset, they perform poorly in the high-dimensional setting of DIV2K and LIDC. Conversely, the proposed equivariant bootstrap algorithm delivers remarkably accurate UQ results even in large-scale problems. ULA tends to provide confidence regions that are only accurate at a specific confidence level, and fail at other levels, whereas diffusion-based methods provide over-confident confidence regions that severely underestimate the uncertainty in the solution. The naive bootstrap also severely underestimates the uncertainty, as discussed in Section 3.

Remarkably, the proposed method, applied with a supervised reconstruction network, simultaneously achieves the most accurate uncertainty quantification results across all tasks as well as the highest image estimation accuracy. The proposed approach also delivers accurate uncertainty estimates when used with an unsupervised network that has been trained from observed data alone, thus removing the need for extensive ground-truth data (we only require a small calibration dataset to set hyperparameters $\sigma_{\theta}$ and $\Delta t$). 

Diffusion-based methods provide the most accurate point estimation results in the inpainting problem. This is related to the fact that the denoisers underpinning the diffusion method were trained on large datasets of natural images which are similar to those in the DIV2K dataset. However, their performance deteriorates in domain-specific images such as tomography scans, even after fine-tuning. For the tomography task, we fine-tuned the DRUNet denoiser~\cite{zhang2017learning} used by diffPIR on the LIDC training set, resulting in an improvement of 0.3 dB in test PSNR (note that \Cref{tab: psnr} shows the performance after fine-tuning). 

\begin{table}[h]
\centering
\caption{Bootstrap parameters.} \label{tab: bootstrap params}
\begin{tabular}{l|ll}
 & Unsup. model & Sup. model \\ \hline
C. Sensing & \textbf{\begin{tabular}[c]{@{}c@{}}$\sigma_{\theta} = 4$\\ $\Delta t =3 $\end{tabular}} & \textbf{\begin{tabular}[c]{@{}c@{}}$\sigma_{\theta} = 0$\\ $\Delta t =3 $\end{tabular}} \\ \hline
Inpainting & \textbf{\begin{tabular}[c]{@{}c@{}}$\sigma_{\theta} = 5$\\ $\Delta t = 10$\end{tabular}} & \textbf{\begin{tabular}[c]{@{}c@{}}$\sigma_{\theta} = 5$\\ $\Delta t = 10$\end{tabular}} \\ \hline
Tomography & \textbf{\begin{tabular}[c]{@{}c@{}}$\sigma_{\theta} = 10 $\\ $\Delta t = 5 $\end{tabular}} &\textbf{\begin{tabular}[c]{@{}c@{}}$\sigma_{\theta} = 8$\\ $\Delta t = 0$\end{tabular}}
\end{tabular}
\end{table}

\begin{figure*}[t]
    \centering
    \includegraphics[width=1\textwidth]{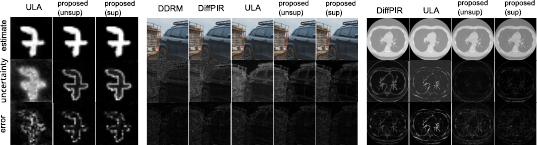}
    \caption{Per-pixel estimation of the error by the evaluated methods. The last row shows the true estimation error. The reader can zoom in to observe the details of the uncertainty maps.} \label{fig: maps all}
\end{figure*}

Lastly, \Cref{fig: maps all} show the reconstructions, the estimation of the marginal per-pixel error (as calculated by the standard deviation), and the true absolute error for each of the evaluated methods. In order to avoid boundary issues in the estimation of the per-pixel errors, we restricted the rotations in the equivariant bootstrap method to multiples of 90 degrees, which have an exact inverse. We observe that the per-pixel error estimates\footnote{This uncertainty visualization analysis can be easily conducted in a multi-resolution manner (see, e.g., \cite{laumont2022bayesian}).} produced by the proposed approach are in good agreement with the true per-pixel errors, whereas the competing methods deliver per-pixel error estimates that are less accurate. 


\begin{figure}[h]
    \centering
    \includegraphics[width=.5\columnwidth]{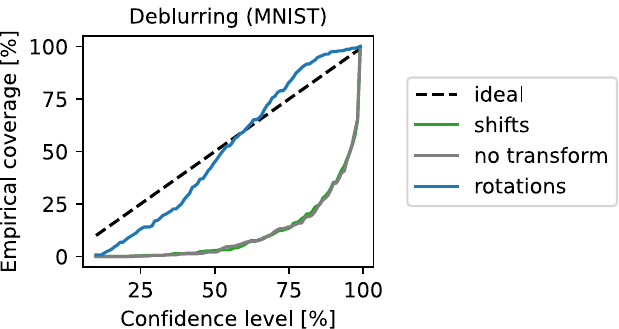}
    \caption{Coverage plots for the anisotropic deblurring problem obtained by bootstrapping with shifts only, rotations only, and no transformations (noise only).} 
    \label{fig: blur}
\end{figure}

\paragraph{Role of Equivariance} We now present a final experiment designed to demonstrate the role of the group action in the uncertainty estimates. We consider a deblurring inverse problem with a highly anisotropic blur kernel, using the MNIST dataset (see \Cref{app: architecture details} for more details). We use the equivariant bootstrap with an estimator $\hat{x}$ that is implemented by using a fully convolutional network, which is by construction shift-equivariant and is trained in a supervised fashion. \Cref{fig: blur} shows the coverage obtained by the proposed equivariant bootstrap technique when $\G$ includes: i) shifts only ($\Delta t=5$ and $\sigma_\theta=0$), and ii) rotations only ($\Delta t=0$ and $\sigma_\theta=5$). As suggested by the analysis in~\Cref{subsec: analysis}, the bootstrap estimates obtained with the shifts obtain the same performance as the naive bootstrap approach, thus under-estimating the true error, as the composition of the forward operator and the reconstruction network is $\G$-equivariant. Conversely, using rotations provides excellent UQ results.

\section{Limitations and future extensions}
The performance of the proposed method is sensitive to the choice of group action. This involves selecting the type of group action (e.g., rotation, shifts), but also choosing the specific actions that will be included in $\G$ (e.g., angles, magnitude of shifts). This calibration can be done on a small evaluation set. However, in some applications, obtaining a reliable evaluation set might be difficult. We leave for future work the study of automatic calibration techniques that can completely bypass the need for ground-truth data.

The proposed method also requires knowledge of the forward operator and noise distribution, which might be mildly misspecified or not fully known in some applications. We leave the analysis of the impact of model misspecification for future work. Future work could also explore blind and semi-blind variants of the proposed equivariant bootstrap.

\section{Conclusion}
Uncertainty quantification is a critical missing component in the modern computational imaging toolbox. Accurately quantifying the uncertainty in restored images is a highly challenging task because of the dimensionality involved and because uncertainty estimates depend strongly on tail probabilities and higher-order moments that are difficult to estimate precisely. The proposed equivariant bootstrap method offers a conceptually simple, yet highly effective strategy for obtaining accurate uncertainty quantification results in imaging inverse problems. The method is also very computationally efficient and scales seamlessly to large settings. 

In addition, the method does not require ground-truth data or a very small amount for calibration tests. When combined with modern unsupervised learning techniques that require measurement data only for training, this enables, for the first time to the best of our knowledge, to perform image point estimation and uncertainty quantification reliably in a fully unsupervised manner. This potent combination is of paramount importance to a wide range of impactful application domains. In particular, applications in science and medicine, where uncertainty quantification is critically important and where obtaining reliable ground-truth data can be extremely expensive or even impossible.

\subsubsection*{Acknowledgements}
This work was supported by the French National Research Agency (ANR) through grant UNLIP and by the UK Research and Innovation (UKRI) Engineering and Physical Sciences Research Council (EPSRC) through grants BLOOM (EP/V006134/1) and LEXCI (EP/W007681/1).

\bibliography{bibliography}

\appendix
\section{Detailed Proofs (Section 4.2)}  \label{app: proofs}

We begin by providing a detailed derivation of the following equality shown in Section 4.2 of the main paper:
\begin{equation*}
\mathbb{E}_{\hat{Y}}\|\tilde{x}(\tilde{Y})-Bx_\star\|_2^2  = \| Rx_\star \|^2 - \| Rx_\star \|_{B_{\star}}^2   + x_\star^\top B^\top  \Pi_{\G}\left( R^\top R + R^{\top}B_\star +B_\star R -2 R \right)Bx_\star 
\end{equation*}
Starting with the definition of the expectation of the equivariant bootstrap error, we have 
\begin{align}
\mathbb{E}_{\hat{Y}}\|\tilde{x}(\tilde{Y})-Bx_\star\|_2^2 &=\frac{1}{{|\G|}}\sum_{g\in\G} \|T^{-1}_g B T_g B x_\star - Bx_\star\|_2^2 \\
&= \frac{1}{{|\G|}}\sum_{g\in\G} \left[\, x_\star^\top B^\top (I_n + T^{-1}_g B^\top B T_g - 2 T^{-1}_g B T_g)Bx_\star\right]\, \\
&= x_\star^\top B^\top \left(I_n + \Pi_{\G}(B^\top B - 2B)\right)Bx_\star\, \label{eq: bstrap error}
\end{align}
Using the decomposition $B=B_{\star}+R$, we have that
\begin{align*}
    \Pi_{\G}(B^\top B -2B) &= \Pi_{\G}(B_{\star}^\top B_{\star} + B_{\star}^\top R + R^\top B_{\star} + R^{\top}R - 2 B_{\star} - 2R) \\
    &= \Pi_{\G}(B_{\star}^\top R + R^\top B_{\star} + R^{\top}R - B_{\star} - 2R) \\ 
    &=  - \Pi_{\G}(B_{\star}) +  \Pi_{\G}(B_{\star}^\top R + R^\top B_{\star} + R^{\top}R - 2R) \\
    &=  - B_{\star} +  \Pi_{\G}(B_{\star}^\top R + R^\top B_{\star} + R^{\top}R - 2R) 
\end{align*}

where we used that $B_{\star}^{\top}B_{\star}=B_{\star}$ and $\Pi_{\G} (B_{\star})= B_{\star}$. Plugging-in this result into \Cref{eq: bstrap error}, we obtain
\begin{align*}
\mathbb{E}_{\hat{Y}}\|\tilde{x}(\tilde{Y})-Bx_\star\|_2^2
&= x_\star^\top B (I-B_{\star}) B x +  x_\star^\top B^\top  \Pi_{\G}\left(B_{\star}^\top R + R^\top B_{\star} + R^{\top}R - 2R\right)Bx_\star\, ,
\end{align*}
We can conclude our derivation by showing that the first term can be written as
\begin{align}
   x_\star^\top B (I-B_{\star}) B x &=  (R x_\star + x_\star)^{\top}  (I-B_{\star}) (R x_\star + x_\star)  \\   
     &= (R x_\star)^{\top} R x_\star - (R x_\star)^{\top} B_{\star} R x_\star + \underbrace{\|x_\star\|^2 - \|B_{\star} x_\star\|^2}_{=0}  + \underbrace{2x_\star^\top R^{\top}x_\star - 2x_\star^\top R^{\top}B_{\star} x_\star}_{=0}  \\
  &= \| Rx_\star \|^2 - \| Rx_\star \|_{B_{\star}}^2 
\end{align}

We now show that an estimator $\hat{x}(\cdot)$ verifying $A\hat{x}(y)=y$ implies that $\| Rx_\star \|_{B_{\star}}^2=0$, which is used in the analysis of the equivariant bootstrap with a linear estimator in Section 4.2 of the main paper.

Using that $\hat{x}(y)=Bx_{\star}$ and $y=Ax_{\star}$, we have that
\begin{align}
A\hat{x}(y)=y \\
ABx_{\star}=Ax_{\star} \\
A(Bx_{\star}-x_{\star}) = 0 
\end{align}
Applying the decomposition $B=B_{\star}+R$ where $B_{\star}x_{\star}=x_{\star}$, we obtain
\begin{align}
A\left((B_{\star}+R)x_{\star}-x_{\star}\right) = 0 \\
ARx_{\star} = 0
\end{align}
Thus, writing $B_{\star}=M_{\star}A$ where $M_{\star}\in\mathbb{R}^{n\times n}$ is the oracle linear estimator, we have 
\begin{align}
    M_{\star}ARx_{\star} = 0\\
    B_{\star} Rx_{\star} = 0
\end{align}
and consequently $\| Rx_\star \|_{B_{\star}}^2=0$ since $B_{\star}$ is positive semi-definite.

\section{Architecture and Training Details} \label{app: architecture details}

\begin{enumerate}
    \item ULA: We use the DnCNN denoiser introduced by~\cite{terris2020building} which is trained on the dataset in~\citep{zhang2017learning}, with a dedicated loss that forces the network to be firmly non-expansive.
    \item DiffPIR: For the tomography problem (single channel images), we finetune the (pretrained) DRUNet denoiser~\citep{zhang2017learning} on the LIDC train set. For the inpainting task (RGB images), we use the attention-based denoiser introduced by~\cite{choi2021ilvr} trained on the FFHQ dataset. 
    \item DDRM: We use the same denoiser as diffPIR~\citep{choi2021ilvr} for the image inpainting task. We didn't evaluate the DDRM method on the compressed sensing and tomography problems since the algorithm can only be applied to problems with a simple singular value decomposition of the forward operator.
    \item Reconstruction networks: 
    \begin{enumerate}
        \item Compressed Sensing (MNIST): For this task, we use an architecture based on an unrolled proximal gradient algorithm with 4 iterations, where the proximal operator was replaced by a trainable U-Net architecture with 4 scales. The same proximal operator network is applied at each iteration (i.e., weight-tied). The resulting unrolled network has 2065033 trainable parameters. The unsupervised network is trained using the robust EI loss~\citep{chen2021robust} with random shifts as transformations. 
        \item Inpainting (DIV2K): We use an architecture based on an unrolled proximal half-quadratic splitting algorithm with 3 iterations, where the proximal operator was replaced by a trainable U-Net architecture with 4 scales. The same proximal operator network is applied at each iteration (i.e., weight-tied). The resulting unrolled network has 8554375 trainable parameters.  The unsupervised network is trained using the robust EI loss~\citep{chen2021robust} with random shifts as transformations.
        \item Tomography (LIDC-IDRI): We use a U-Net architecture applied to the filtered-back projected images, similarly to~\cite{jin2017deep}. The unsupervised network is trained using the robust EI loss~\citep{chen2021robust} with random rotations as transformations. The network has 8553088 trainable parameters.
        \item Anistropping deblurring (MNIST): We choose a blur forward operator with a vertical $7\times 1$ kernel with entries $\frac{1}{7}$ (i.e., summing to one).
        We use an architecture based on an unrolled proximal gradient algorithm with 3 iterations, where the proximal operator was replaced by a DnCNN architecture~\citep{zhang2017beyond} with circular padding. The same proximal operator network is applied at each iteration (i.e., weight-tied). The resulting unrolled network has 664711 trainable parameters and is fully shift-equivariant (see~\citep{celledoni2021equivariant} for more details on unrolled equivariant networks).
    \end{enumerate}
     The unrolled architectures are generated using the DeepInverse library~\citep{deepinverse}. We use the same architecture for supervised and unsupervised learning.  All the U-Net architectures used for the supervised and unsupervised reconstruction networks are based on the backbone in~\citep{chen2021equivariant}. All networks are trained using the Adam optimizer with standard hyper-parameters ($\beta_1=0.9$ and $\beta_2=0.999$) and a step size of $10^{-4}$.

\end{enumerate}

\section{Additional Results} \label{app: additional results}
\begin{figure}[h]
    \centering
    \includegraphics[width=.6\textwidth]{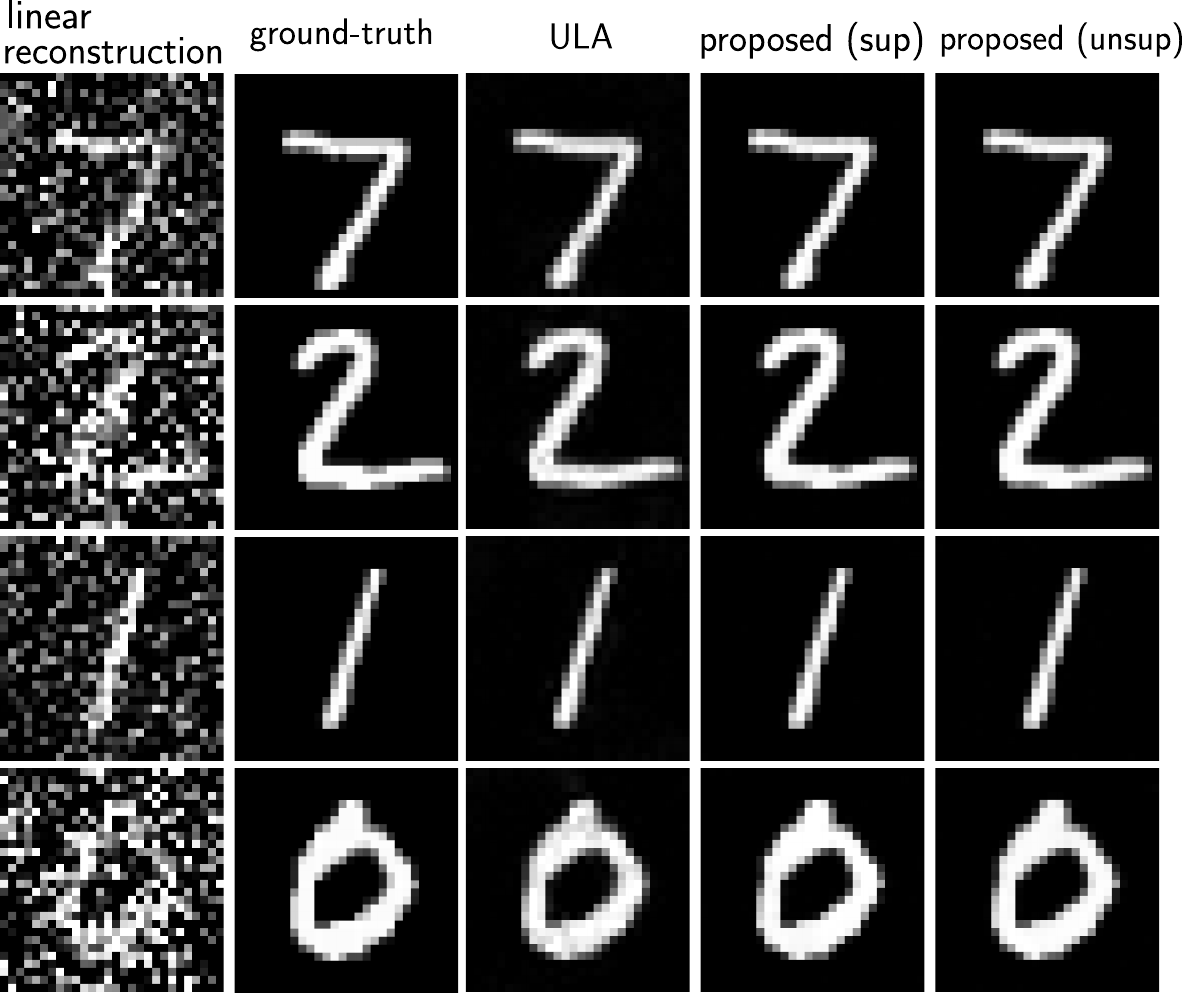}
    \caption{Reconstructed test images by the different evaluated methods for the compressed sensing task using MNIST. The first column shows a simple a simple linear reconstruction of the measurements, i.e., $\hat{x}(y)=A^{\top}y$.} \label{fig: mnist}
\end{figure}

\begin{figure}[h]
    \centering
    \includegraphics[width=.8\textwidth]{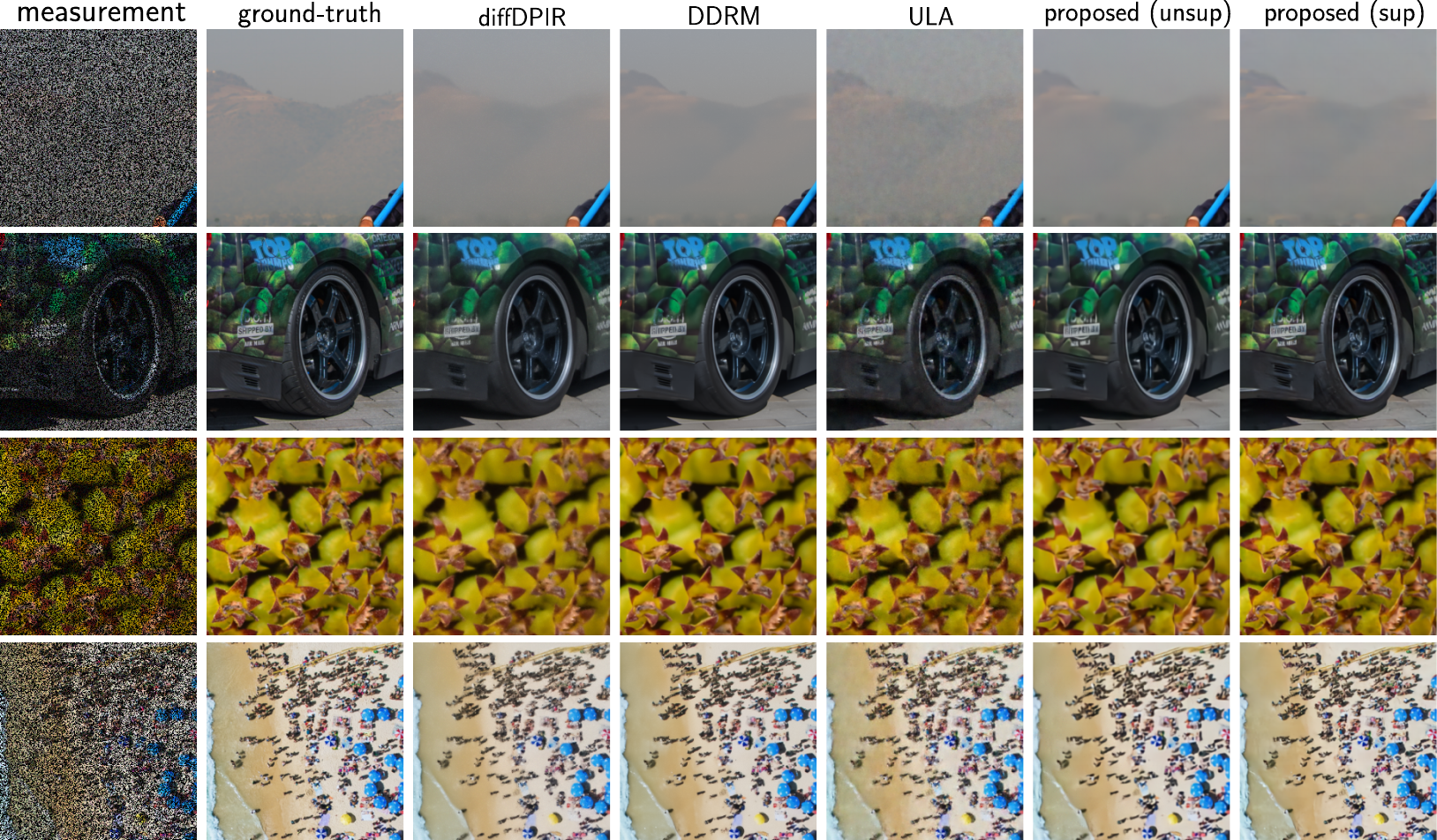}
    \caption{Reconstructed test images by the different evaluated methods for the sparse angle image inpainting task using the DIV2K dataset.} \label{fig: inpainting}
\end{figure}

\begin{figure}[h]
    \centering
    \includegraphics[width=.8\textwidth]{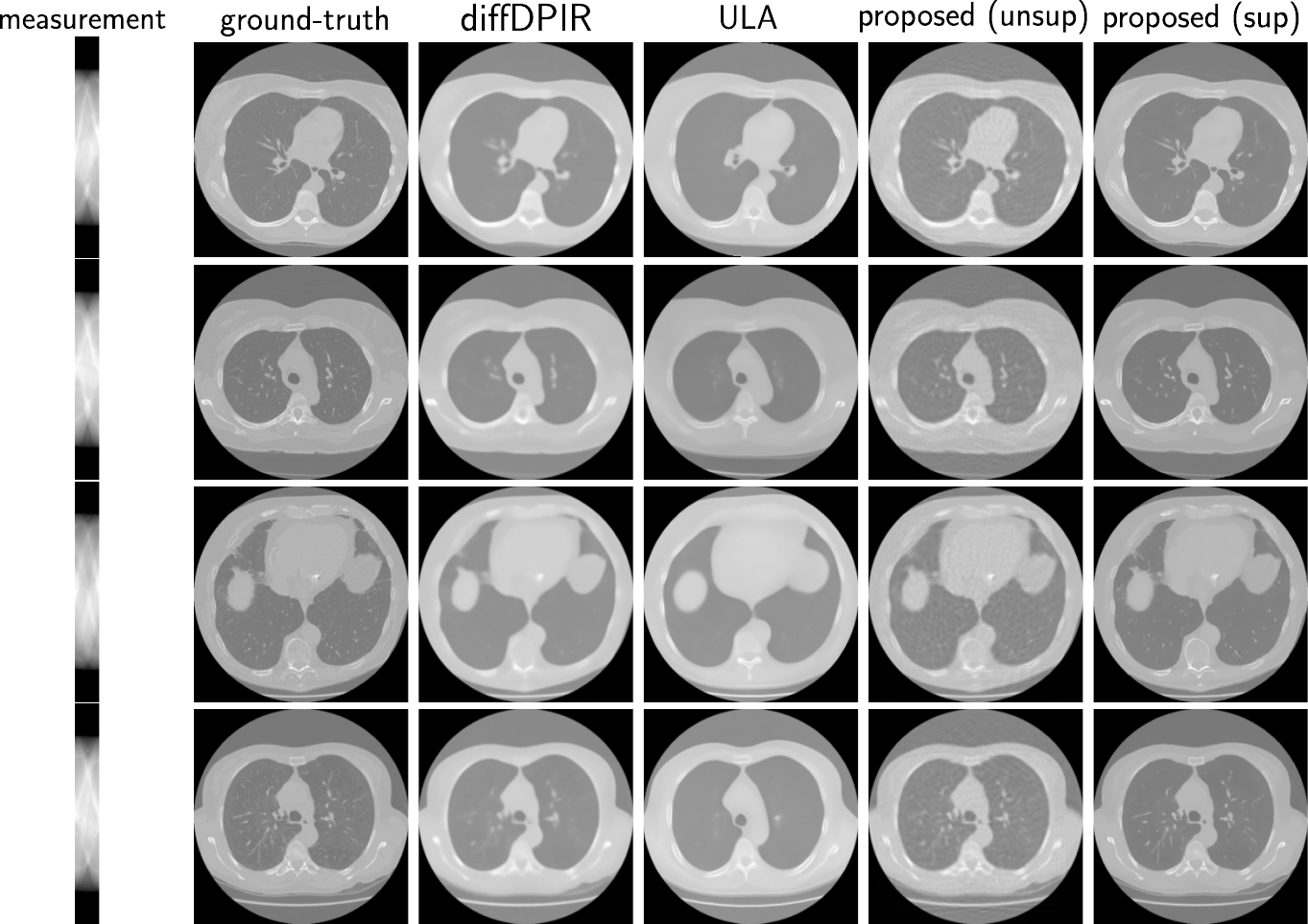}
    \caption{Reconstructed test images by the different evaluated methods for the sparse angle tomography task using the LIDC-IDRI dataset. The first column shows the sinograms $y$.} \label{fig: CT}
\end{figure}

\Cref{fig: mnist,fig: inpainting,fig: CT} show test reconstructions obtained by the different evaluated methods for the problems of compressed sensing, image inpainting, and sparse-angle tomography.

\end{document}